\documentclass[twocolumn,prl,showpacs]{revtex4}%
\usepackage{graphicx}%
\usepackage{amsmath}%
\usepackage{amsfonts}%
\usepackage{amssymb}

\def\d{{\partial}}

\def\k{{ {\bf k} }}
\def\p{{ {\bf p} }}
\def\q{{ {\bf q} }}
\def\Q{{ {\bf Q} }}

\def\w{{\omega}}
\def\a{{\alpha}}
\def\b{{\beta}}

\begin{document}
\title{
%Strong impurity pair-breaking effect on sign-reversing \\ 
%fully-gapped superconducting state in iron pnictides
Violation of Anderson's Theorem for Sign-Reversing $s$-Wave \\
State of Iron-Pnictide Superconductors
}
\author{Seiichiro \textsc{Onari}$^{1}$,
and Hiroshi \textsc{Kontani}$^{2}$}
\date{\today }

\begin{abstract}
%To understand the pairing symmetry in iron pnictides,
%observed robustness of superconductivity against impurities 
%offer us significant information.
%To understand the pairing symmetry in iron pnictides,
%impurity effect is expected to offer us significant information.
%To understand the essential nature of the multiorbital superconductivity 
%in iron pnictides,
Based on the five-orbital model,
we study the effect of local impurity in iron pnictides, 
%using the $T$-matrix approximation,
and find that the interband impurity scattering is promoted by
the $d$-orbital degree of freedom.
This fact means that the fully-gapped sign-reversing $s$-wave state, 
which is predicted by spin fluctuation theories,
is very fragile against impurities.
%This is a natural consequence of the existence of orbital degree of freedom 
%in the five-orbital model.
%basic structure of the $d$-orbital Bloch function in the five-orbital model.
%owing to the momentum dependence of $d$-orbital Bloch function.
% is treated correctly.
%the two kinds of $s$-wave superconducting states.
%In the case of 
In the BCS theory, only 1\% impurities with intermediate strength
induce huge pair-breaking, resulting in the large in-gap state 
and prominent reduction in $T_{\rm c}$,
contrary to the prediction based on simple orbital-less models.
%models without orbital degree of freedom.
%, as in the case of non-$s$-wave SC state.
%the orbital degree of freedom is important not only 
%on the impurity effect in multiorbital superconductors.
%This prominent impurity effect on the $s_\pm$ state disappears erroneously
%if we 
%treat the multiorbital bandstructure incorrectly.
%The present analysis suggests the absence of sign reversals 
%in the gap in iron pnictides.
The present study provides a stringent constraint
on the pairing symmetry and the electronic states in iron pnictides.
%on the superconducting electronic properties.
\end{abstract}
%\draft

\address{
$^1$ Department of Applied Physics, Nagoya University and JST, TRIP, 
Furo-cho, Nagoya 464-8602, Japan. 
\\
$^2$ Department of Physics, Nagoya University and JST, TRIP, 
Furo-cho, Nagoya 464-8602, Japan. 
%$^3$ 
%JST, Transformative Research-Project on Iron Pnictides (TRIP), 
%Chiyoda, Tokyo 102-0075, Japan.
%JST, TRIP, Nagoya University, Furo-cho, Nagoya 464-8602, Japan. 
}
 

\sloppy

\maketitle

%%%%%%%%%%%%%%%%%%
%Introduction
%%%%%%%%%%%%%%%%%%

Since the discovery of 
high-$T_{\rm c}$ superconductors with FeAs layers 
 \cite{Hosono},
the symmetry of the superconducting (SC) gap 
has been studied very intensively.
NMR studies had revealed that 
the singlet SC state is realized in iron pnictides 
 \cite{Kawabata,Grafe,Zheng}.
The realized gap function is isotropic and band-dependent 
according to the penetration depth measurement \cite{Matsuda}
and the angle-resolved photoemission spectroscopy (ARPES)
 \cite{ARPES1,ARPES4}.
This result is contrasting to high-$T_{\rm c}$ cuprates,
where the nodal $d$-wave state is realized.
The fully-gapped state is also supported by 
the rapid suppression in $1/T_1$ ($\propto T^{n}$; $n\sim6$) 
below $T_{\rm c}$ in several compounds \cite{Sato-T1}.
On the other hand, the relation $1/T_1 \propto T^3$ had been reported 
in different compounds \cite{Ishida,Mukuda}.
%which might suggest highly anisotropic or nodal gap state.

\begin{figure}[!htb]
\includegraphics[width=.5\linewidth]{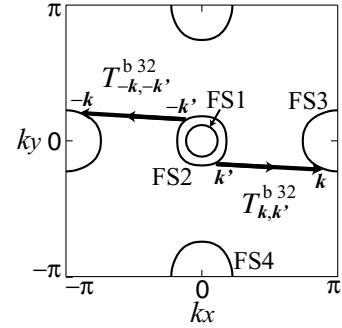}
\caption{
Hole-pockets (FS1,2) and electron-pockets (FS3,4) in iron pnictides.
$|T_{\k,\k'}^{3,2}|^2$ represents the impurity induced 
pair-hopping amplitude between (FS2, $\pm\k'$) and (FS3, $\pm\k$).
%both of which are mainly composed of $xz$ and $yz$ orbitals.
%Pair hoppings between hole and electron-pockets 
%destroy the $s_{\pm}$-wave state.
}
\label{fig:FS}
\end{figure}

Figure \ref{fig:FS} shows
the Fermi surfaces (FSs) in the unfolded Brillouin zone 
\cite{Kuroki}.
%for 10\% electron-doped case, i.e., the electron number is 6.1 per Fe ion.
Hereafter, we fix the electron number as 6.1 (10\% electron-doped case).
%The density of states (DOS) for FS1,4 and FS2,3 are 
%$N_{1+4}(0)= 0.37$ and $N_{2+3}(0)=0.29$ per spin at $\mu$.
Then, the total density of states (DOS) per spin is 
$N(0)= 0.66$ ${\rm eV}^{-1}$ at the Fermi level.
In the presence of the Coulomb interaction, antiferromagnetic (AF) 
fluctuations with $\Q\approx (\pi,0)$ is expected to emerge
due to the nesting between FS1,2 and FS3,4
 \cite{Mazin}.
Based on this fact, a fully gapped $s$-wave state with sign reversal 
($\Delta_{1,2}>0$, $\Delta_{3,4}<0$),
which is called the $s_\pm$-wave state,
had been predicted theoretically  \cite{Kuroki,Mazin}.
% \cite{Kuroki,Mazin,RG,Nomura,Yanagi,Ikeda}.
%Also, possibility of a conventional $s$-wave state without sign reversal
%($s_{++}$-wave state) due to phonon mechanism
%had been discussed in Ref. \cite{EP}.
However, no conclusive evidence for the $s_\pm$-wave state 
has been obtained experimentally so far.

%In the history of studying the superconductivity,
Historically, the study of 
impurity effects had offered us significant information in 
determining the pairing symmetry in many superconductors. 
In iron pnictides, the SC state survives against high substitution of
Fe sites by other element (more than 10\%), like Co, Rh, Ni, Zn, Ru, and Ir
 \cite{Kawabata,Co,Rh,Ni,Zn,Ru,Ir,122-phase,Canfield}.
%Recently, the phase diagram for electron-doped system
%Ba(Fe$_{1-x}$Co$_x$)$_2$O$_2$ had been established \cite{122-phase}.
In Ba(Fe$_{1-x}$Co$_x$)$_2$As$_2$ \cite{122-phase,Canfield},
the AF ordered state is removed at $x=0.07$, and SC state 
with $T_{\rm c}=24$ K appears.
$T_{\rm c}$ gradually decreases as $x$ increases, and 
it reaches zero at $x=0.17$.
This phase diagram suggests the relation
$-\Delta T_{\rm c}\sim 2$K per 1\% Co substitution,
and cannot be understood if the impurity pair-breaking
gives $-\Delta T_{\rm c}> 20$ K/\% like in high-$T_{\rm c}$ cuprates.
% \cite{Sato-private}.
According to the first principle calculation, the impurity 
potential due to Co substitution for $xz,yz$-orbitals is 1.52 eV,
and its radius is only 1 ${\buildrel _{\circ} \over {\mathrm{A}}}$
 \cite{impurity-pot}.
Also, a bulk superconducting state with $T_{\rm c}=24$ K is realized 
in Sr(Fe$_{1-x}$Ir$_x$)$_2$As$_2$ for $x\sim0.25$ \cite{Ir}.
These experimental facts would eliminate the possibility of
nodal gap state.

%%%%%%%%%%%%%%%%%%%%%%%%%%%%%%%%%
%%%%%%%%%%%%%%%%%%%%%%%%%%%%%%%%%

To explain these experiments,
simple multiband BCS models with constant
impurity potential $I_{\a,\b}^{\rm b}$ in the band-diagonal basis
($\a,\b$ being the FSs), which we call the 
``orbital-less multiband model'', had been studied intensively.
In the Born regime ($I^{\rm b} \ll (\pi N(0))^{-1}$),
the $s_\pm$-wave state is suppressed by interband scattering 
%of Cooper pair between bands with $\Delta>0$ and $\Delta<0$
 \cite{Chubukov,Parker,Tesanovic}.
In the unitary regime, % where $I^{\rm b}\gg (\pi N(0))^{-1}$,
in contrast, the $s_\pm$-wave state is very robust
since the interband (intraband) scattering is renormalized to zero
(finite) if ${\hat I}^{\rm b}$ is constant and ${\rm det}\{{\hat I}^{\rm b}\}\ne0$
 \cite{Senga,Bang}.
That is, Anderson's theorem \cite{Anderson}
is recovered in the unitary regime.
%and thus experimental robustness of $T_{\rm c}$ 
%had been considered to support the $s_\pm$-wave state.
% since the Fe-site substitution will induce a strong potential.
However, the latter result
should be re-examined since ${\hat I}^{\rm b}$ 
has complex momentum-dependence in usual multiorbital systems.
% however, local impurity potential 
%has momentum-dependence in the band-diagonal basis.
%but in the $d$-orbital basis.
%This fact had not been considered seriously.

In this letter, we present the first study of the impurity effect 
on the bulk DOS and $T_{\rm c}$ based on the five-orbital model
given in Ref. \cite{Kuroki}.
%for both $s_\pm$- and $_{++}$-wave states.
%In view of the Fe-site substitution, 
%momentum-dependence of the local impurity potential due to 
By treating the multiorbital effect correctly, 
we reveal that the Anderson's theorem is seriously violated for 
the $s_\pm$-wave state, %in the five-orbital model,
due to the interband scattering in Fig. \ref{fig:FS},
mainly via $xz$ and $yz$ orbitals.
For this reason, only 1\% impurities with a moderate or strong 
potential induce large in-gap DOS and prominent reduction in $T_{\rm c}$
($-\Delta T_{\rm c}\sim 20$ K/\%) in the BCS theory,
which are comparable to those in nodal gap SC states.
%comment
The present study suggests a reasonable possibility that a
conventional $s$-wave state without sign reversal
($s_{++}$-wave state) would be realized in dirty iron pnictides.

%In frequently used orbital-less multiband models, in contrast, 
%Anderson's theorem holds for the $s_\pm$-wave state
%in the unitary limit, unless det$\{{\hat I}^{\rm b}\}=0$ \cite{Senga}.
%contrary to the prediction based on orbital-less models
%for det$\{{\hat I}^{\rm b}\}\ne0$ \cite{Senga}.
%Since the interband pair-scattering leads to the violation 
%of Anderson's theorem in the $s_\pm$-wave state, only 1\% impurities 
%inevitably induces large in-gap DOS and prominent reduction 
%in $T_{\rm c}$; $-\Delta T_{\rm c}\sim 30 $K/\%,
%which is comparable to that in nodal gap states.
%This result will be guaranteed 
%independently of the specific pairing mechanism,
%as long as the multiorbital bandstructure 
%in the five-orbital model in Ref. \cite{Kuroki} is adequate.
%a direct consequence of the fact that 
%the relative weight of five $d$-orbitals varies with $\k$ on each FS,
%independently of detailed SC gap structure and so on.
%the multiorbital nature of
%the Bloch electrons described by the five-orbital model, 
%The obtained result well (but accidentally)
%corresponds to the special case det$\{{\hat I}^{\rm b}\}=0$
%in the orbital-less model \cite{Senga,Bang}.

%Thus, previous analysis of unitary impurity effect
%based on orbital-less models are erroneous for iron pnictides.

%As a result, it is difficult to explain the smallness of
%the impurity effect in iron pnictide superconductors
%in terms of the $s_\pm$-wave state.

%MODEL
Hereafter, we study the following 
$10\times10$ Nambu BCS Hamiltonian 
in the real $d$-orbital basis; $t_{\rm 2g}$ ($xy$, $yz$, $zx$) 
and  $e_g$ ($x^2\mbox{-}y^2$, $z^2$):
\begin{eqnarray}
{\hat {\cal H}}_\k=\left(
\begin{array}{cc}
{\hat H}_\k^0 & {\hat \Delta}_\k \\
{\hat \Delta}_\k^\dagger & -{\hat H}_{\k}^0 \\
\end{array}
\right) ,
 \label{eqn:H0}
\end{eqnarray}
where ${\hat H}_\k^0$ is the $5\times5$ hopping matrix of the five-orbital 
tight-binding model, which was introduced in Ref. \cite{Kuroki}.
%comment
$\Delta_\k^{j,l}= \sum_\a U_\k^{j,\a} \Delta_{\k,\a} {U_\k^{l,\a}}^*$
is the singlet gap function in the $d$-orbital basis,
where $\Delta_{\k,\a}$ is the gap function for FS$\a$
in the band-diagonal basis, 
and $U_\k^{j,\a}= \langle \k;j|\k;\a\rangle$ is the $5\times5$
transformation unitary matrix between two representations.
Then, the Green function is given by
\begin{eqnarray}
{\hat {\cal G}}_\k(i\w_n) &\equiv&
\left(
\begin{array}{cc}
{\hat G}_\k(i\w_n) & {\hat F}_\k(i\w_n) \\
{\hat F}^\dagger_\k(i\w_n) & -{\hat G}_{\k}(-i\w_n) \\
\end{array}
\right)^{-1} 
 \nonumber \\
&=& (i\w_n{\hat 1}-{\hat \Sigma}_\k(i\w_n)-{\hat {\cal H}}_\k)^{-1} ,
 \label{eqn:G}
\end{eqnarray}
where $\w_n=\pi T(2n+1)$ is the fermion Matsubara frequency,
${\hat G}_\k$ (${\hat F}_\k$) is the $5\times5$ normal (anomalous)
Green function, and ${\hat \Sigma}_\k$ is the self-energy 
in the $d$-orbital basis.

First of all, we calculate the impurity effect on the DOS in the SC state
using the $T$-matrix approximation, which is reliable 
when the impurity concentration $n_{\rm imp}$ is much smaller than unity.
We consider the local impurity potential 
due to the substitution of Fe by other 3$d$ elements:
%If we neglect the deformation of the As-tetrahedron,
In the present $d$-orbital basis,
the impurity potential is momentum-independent and diagonal.
Then, the $T$-matrix for a single impurity,
which is $\k$-independent in the $d$-orbital basis, is given as
\begin{eqnarray}
{\hat {\cal T}}(i\w_n)= ({\hat 1}-{\hat {\cal I}}
{\hat {\cal G}}_{\rm loc}(i\w_n))^{-1} {\hat {\cal I}},
 \label{eqn:Tmat}
\end{eqnarray}
where ${\hat {\cal G}}_{\rm loc}(i\w_n)
\equiv\frac1N \sum_\k {\hat {\cal G}}_\k(i\w_n)$.
Neglecting the crystalline-field splitting between $t_{\rm 2g}$ and 
$e_{\rm g}$, the impurity potential ${\hat {\cal I}}$ is simply given as
${\cal I}_{j,l}= I\delta_{j,l}$ for $1\le j \le5$, and 
${\cal I}_{j,l}= -I\delta_{j,l}$ for $6\le j \le10$.
In the $T$-matrix approximation, the self-energy matrix
in the $d$-orbital basis is $\k$-independent.
It is given as
\begin{eqnarray}
{\hat \Sigma}(i\w_n) &\equiv&
%\left(
%\begin{array}{cc}
%\delta{\hat \Sigma}^n(i\w_n) & \delta{\hat \Sigma}^a(i\w_n) \\
%\delta{\hat \Sigma}^a(i\w_n) & -\delta{\hat \Sigma}^n(-i\w_n) \\
%\end{array}
%\right) 
% \nonumber &=& 
n_{\rm imp}{\hat {\cal T}}(i\w_n) .
 \label{eqn:Self}
\end{eqnarray}
%
%where $\delta{\hat \Sigma}^{n(a)}$ is the normal (anomalous) self-energy.

The gap function ${\hat \Delta}_\k$ in eq. (\ref{eqn:H0}) 
is given by the solution of the Eliashberg equation:
\begin{eqnarray}
\Delta_\k^{j,j'}(i\w_n)
= -\frac{T}{N}\sum_{\q,m} \sum_{l,l'}V_{\k,\q}^{j,j';l,l'}(i\w_n,i\w_m)
F_\q^{l,l'}(i\w_m) ,
 \label{eqn:Eliash1}
\end{eqnarray}
where $V_{\k,\q}^{j,j';l,l'}$ is the pairing potential.
In the fully self-consistent $T$-matrix approximation,
we solve eqs. (\ref{eqn:G})-(\ref{eqn:Eliash1}) self-consistently.
%comment
In calculating the DOS, however, we solve only 
eqs. (\ref{eqn:G})-(\ref{eqn:Self}) self-consistently, 
assuming the isotropic $\Delta_\a$ for FS$1\sim4$.
%We apply this partially self-consistent calculation since
%we are interested in universal properties that are
%independent of the detail of the pairing potential.
As we will show below, 
the reduction in $T_{\rm c}$ in the $s_\pm$-wave state
at $n_{\rm imp}=0.01$ exceeds 20 K for $I\gg W_{\rm band}$.
Then, the reduction in the SC gap is $\sim 36$ K
if the relation $\Delta/T_{\rm c}=1.8$ is supposed.
Hereafter, we assume that this reduction in ${\hat \Delta}$ 
had been included from the beginning.

%We do not take this reduction in $|\Delta_\a|$
%assuming that the reduction of $|\Delta_\a|$ due to pair-breaking 
%had been included in from the beginning.
%we consider that the reduction of ${\hat \Delta}$ 
%due to impurity pair-breaking had been included from the beginning.

%%%%%%%%%%%%%%%%%%%%%%%%%%%%%%%%%
\begin{figure}[!htb]
\includegraphics[width=.75\linewidth]{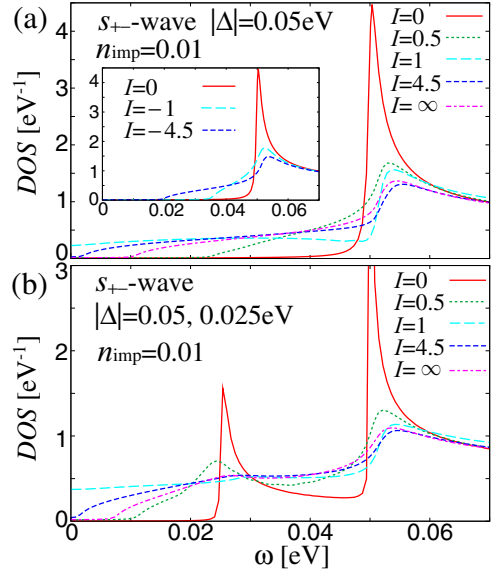}
\caption{(color online)
Obtained DOS in the $s_\pm$-wave state for $n_{\rm imp}=0.01$.
(a) $|\Delta_{1,2,3,4}|=0.05$ eV, and
(b) $|\Delta_{1,3,4}|=2|\Delta_2|=0.05$ eV.
%(a) $\Delta_{1,2}=-\Delta_{3,4}=0.05$ eV, and
%(b) $\Delta_1=2\Delta_2=-\Delta_{3,4}=0.05$ eV.
}
\label{fig:DOS-s-pm}
\end{figure}
%%%%%%%%%%%%%%%%%%%%%%%%%%%%%%%%%
Figure \ref{fig:DOS-s-pm} shows the 
obtained DOS in the $s_\pm$-wave SC state per spin
for $n_{\rm imp}=0.01$ at $T=0$,
assuming the isotropic gap in the band-diagonal basis.
The number of $\k$-meshes is $N=512\times512$.
The DOS is given by the normal Green function in eq. (\ref{eqn:G}) as
$N(\w)=\frac1{\pi N} \sum_{\k} {\rm Im}\{{\rm Tr}{\hat G}_\k(\w-i\delta)\}$.
In Fig \ref{fig:DOS-s-pm} (a), we put 
$\Delta_{1,2}=-\Delta_{3,4}=0.05$ eV.
We also study the case where only $\Delta_2$ is changed to $0.025$ eV
in Fig \ref{fig:DOS-s-pm} (b),
consistently with ARPES measurements
 \cite{ARPES1,ARPES4}. 
Since these gaps are a few times larger than experimental values,
the obtained impurity effect is underestimated.
%, as we discuss later.
%The gap function in the $d$-orbital basis in eq. (\ref{eqn:H0}) becomes
%$\Delta_\k^{j,l}= \sum_\a U_\k^{j,\a} {\Delta_\a U_\k^{l,\a}}^*$,
%where $U_\k^{j,\a}= \langle \k;j|\k;\a\rangle$ is the $5\times5$
%transformation unitary matrix between two representations.

In Fig. \ref{fig:DOS-s-pm}, prominent in-gap state 
due to impurity pair-breaking emerges even for small value of $I=0.5$,
which means that only 1\% impurity concentration induces 
strong pair-breaking effect in the $s_\pm$-wave state.
This result is inconsistent with the analysis
in the orbital-less model, where pair-breaking due to 
unitary impurity is absent in the $s_\pm$-wave state.
Now, we explain that this discrepancy
arises from the presence or absence of orbital degree of freedom:
In the band-diagonal basis, the %$5\times5$ 
$T$-matrix in the normal state is given by the solution of
\begin{eqnarray}
{\hat T}_{\k,\q}^{\rm b}= {\hat I}_{\k,\q}^{\rm b}+
\frac1N \sum_{\p} {\hat I}_{\k,\p}^{\rm b} {\hat G}^{\rm b}_\p {\hat T}^{\rm b}_{\p,\q} ,
 \label{eqn:Tb}
\end{eqnarray}
where ${\hat I}^{\rm b}_{\k,\q}= I {\hat U}^\dagger_\k {\hat U}_\q$
is the impurity potential in the band basis.
${\hat G}^{\rm b}_\p$ is the normal Green function that is band-diagonal.
When $\k$-dependence of ${\hat U}_\k$ is small on each FS,
${\hat I}^{\rm b}$ becomes constant by replacing $U^{l,\a}_\k$
with $\langle U^{l,\a}_\k \rangle_{\k\in{\rm FS\a}}$, 
which corresponds to the orbital-less model
 \cite{Chubukov,Parker,Tesanovic,Senga,Bang}.
Using ${\hat g}_{\rm loc}^{\rm b}\equiv \frac1N \sum_\k{\hat G}_\k^{\rm b}$,
the solution is given as
%eq. (\ref{eqn:Tb}) is reduced to
%${\hat T}^{\rm b}= {\hat I}^{\rm b}+
%{\hat I}^{\rm b} {\hat g}^{\rm b}_{\rm loc} {\hat T}^{\rm b}$,
%where ${\hat g}_{\rm loc}^{\rm b}\equiv \frac1N \sum_\k{\hat G}_\k^{\rm b}$
%
\begin{eqnarray}
{\hat T}^{\rm b}= ({\hat 1}- {\hat I}^{\rm b} {\hat g}_{\rm loc}^{\rm b})^{-1} {\hat I}^{\rm b}, 
\label{eqn:Tb2}
\end{eqnarray}
which becomes band-diagonal as 
$T_{\a,\b}^{\rm b} = -1/g_{{\rm loc},\a}^{\rm b}\cdot \delta_{\a,\b}$
when $I\rightarrow \infty$,
as far as ${\rm det}\{{\hat I}^{\rm b}\}\ne0$.
Thus, the pair-breaking due to interband scattering,
which is described in Fig. \ref{fig:FS},
is absent for $I=\infty$ in the orbital-less model.

However, the above discussion does not hold
in iron pnictides since ${\hat U}_\k$ strongly depends on $\k$:
For example, $|U_\k^{zx,1}|^2\sim \cos^2\theta_\k$
and $|U_\k^{yz,1}|^2\sim \sin^2\theta_\k$ for FS1
(where $\theta_\k=\tan^{-1}(k_y/k_x)-\pi/4$),
and $xz,yz$ orbitals also constitute a large part of FS3,4
 \cite{Kuroki}.
%(Here, the $d$-orbital coordinate is 
%rotated from the unfolded zone by $\pi/4$.)
In fact, we have verified numerically that ${\hat T}^{\rm b}_{\k,\p}$ 
has large offdiagonal elements \cite{comment2}.
For this reason, the impurity pair-breaking 
for $s_\pm$-wave state is as large as that in the nodal gap state.
%The prominent local impurity effect in Fig. \ref{fig:DOS-s-pm}
%in the five-orbital model
%would have some relation to recent studies based on two-orbital models
%reporting the existence of a kind of Andreev bound state 
%\cite{ABS1,ABS2,ABS3} 
%and in-gap state \cite{localDOS} in the $s_\pm$-wave SC state.

As recognized in Fig. \ref{fig:DOS-s-pm},
%the impurity effect strongly depends on the sign of $I$ 
%except for $I=\infty$.
the impurity pair-breaking effect strongly depends on %${\rm sgn}\{I\}$ 
the sign of $I$ except for $I=\infty$:
The impurity effect is stronger for $I>0$, 
and it is the most prominent at $I\sim+1.5$eV.
This result originates from the fact that the quasiparticle damping
$\gamma_\a= n_{\rm imp}{\rm Im}T_{\a,\a}^{\rm b}(-i0)$, which works as the 
depairing effect when the Anderson's theorem is not satisfied,
depends on the sign of $I$  %${\rm sgn}\{I\}$ 
in the presence of strong particle-hole asymmetry.
\cite{ROP,comment}.
For the same reason, we find that the residual resistivity
in the $T$-matrix approximation, which is given by
$c/\rho_{\rm imp}=({2e^2}/h)\sum_\k {\rm Tr}
\{ \d{\hat H}_\k^0/\d k_x \cdot {\hat G}_\k(+i0) \cdot
 \d{\hat H}_\k^0/\d k_x \cdot {\hat G}_\k(-i0) \}$
for $\Delta=0$ ($h/e^2=6.45$ k$\Omega$, and
$c \approx 6 {\buildrel _{\circ} \over {\mathrm{A}}}$ 
is the inter-layer spacing) \cite{ROP},
is larger for positive $I$:
$\rho_{\rm imp}(n_{\rm imp}=0.01)=22$, 14, and 10 
[$\mu\Omega {\rm cm}$] at $I=1$, 4.5, and $\infty$ [eV], respectively.
For $I<0$, $\rho_{\rm imp}(n_{\rm imp}=0.01)=3$ and 7
[$\mu\Omega {\rm cm}$] at $I=-1$ and $-4.5$ [eV], respectively.
%For $|I|\ll1$ eV, $\rho_{\rm imp}$ is proportional to $I^2$.
In Ref. \cite{Sato-imp}, $\rho_{\rm imp}$ for 1\% Co doping 
in polycrystalline Nd1111 exceeds 120 $\mu\Omega$cm, 
which corresponds to 30 $\mu\Omega$cm in single crystal.
$\rho_{\rm imp}$ in Co-doped La1111 is still larger.
Near AF quantum-critical-point, $\rho_{\rm imp}$ can exceed the value
given by the $T$-matrix approximation since the many-body correlations 
are enhanced around the impurity, as discussed in Ref. \cite{ROP}.
In under-doped high-$T_{\rm c}$ cuprates, for example,
$\rho_{\rm imp}$ due to 1\% Zn impurity reaches $100\ \mu\Omega$cm,
which is 10 times of the residual resistivity in over-doped cuprates \cite{ROP}.

%from the residual resistivity $n_{\rm imp}$.
%In the $T$-matrix approximation,
%For $I\ll1$ eV,
%$\rho_{\rm imp}(n_{\rm imp}=0.01) \approx 11.2 I^{-2}$ $\mu\Omega {\rm cm}$.
%$n_{\rm imp}^{\rm c} \approx 0.02 I^{-2}$ in case (a),
%Note that $\rho_{\rm imp}$ is independent of $z$.

%%%%%%%%%%%%%%%%%%%%%%%%%%%%%%%%%
\begin{figure}[!htb]
\includegraphics[width=.75\linewidth]{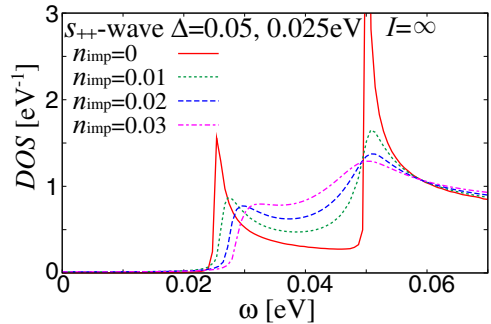}
\caption{(color online)
Obtained DOS in the $s_{++}$-wave state for $I=\infty$.
We put $\Delta_{1,3,4}=2\Delta_2=0.05$ eV.
}
\label{fig:DOS-s-pp}
\end{figure}
%%%%%%%%%%%%%%%%%%%%%%%%%%%%%%%%%
We also study the impurity effect on the $s_{++}$-wave state
with $\Delta_{1,3,4}=0.05$ eV and $\Delta_2=0.025$ eV
in Fig. \ref{fig:DOS-s-pp}.
In this case, pair-breaking due to interband scattering is very small
since the sign of $\Delta_\a$ on all the FSs are the same.
%(i.e., Anderson's theorem).
As $n_{\rm imp}$ increases, the structure of DOS is 
gradually smoothened.
%, suggesting that the Hebel-Schlicter peak becomes tiny or absent.
The absence of zero-energy state is consistent with
many NMR measurements \cite{Kawabata,Grafe,Zheng,Ishida,Mukuda}.
% if we solve fully self-consistently,
% smaller gap will quickly decay, and the anisotropy will increase.

Next, we calculate the impurity effect on $T_{\rm c}$.
To demonstrate the qualitative difference between 
$s_\pm$-wave and $s_{++}$-wave states,
we introduce the following BCS-type 
pairing interaction in the band basis \cite{Allen}:
\begin{eqnarray}
V_{\k,\q}^{\a,\b}(i\w_n,i\w_m)= g_{\a,\b} \cdot 
\phi_\k\phi_\q \cdot \xi(\w_n)\xi(\w_m)
%& &V_{\k,\q}^{j,j';l,l'}(i\w_n,i\w_m)= g_{\k,\q}^{j,j';l,l'} \cdot 
%\phi_\k\phi_\q \cdot \xi(\w_n)\xi(\w_m), \\
%& &g_{\k,\q}^{j,j';l,l'} 
%= \sum_{\a,\b}g_{\a,\b}\cdot U_\k^{j,\a}(U_\k^{j',\a})^* \cdot
% (U_\q^{l,\b})^*U_\q^{l',\b} ,
 \label{eqn:V}
\end{eqnarray}
where $\phi_\k=1$ for $s_{++}$-wave, and
$\phi_\k={\rm sgn}\{\cos k_x \cos k_y\}$ for $s_\pm$-wave.
We also put $\xi(\w_n)=\w_D^2/(\w_n^2+\w_D^2)$,
where $\w_D$ is the cutoff energy.
$V_{\k,\q}^{j,j';l,l'}$ in eq. (\ref{eqn:Eliash1}) is given by
$\sum_{\a,\b}V^{\a,\b}\cdot U_\k^{j,\a}(U_\k^{j',\a})^* 
\cdot (U_\q^{l,\b})^*U_\q^{l',\b}$.
Then, $T_{\rm c}$ is obtained by solving the  
%inserting eq. (\ref{eqn:V})
%into the Eliashberg equation (\ref{eqn:Eliash1}), and solving the 
eqs. (\ref{eqn:G})-(\ref{eqn:Eliash1}) fully self-consistently.
Here, we do not consider the mass-enhancement
since the energy derivative of the self-energy 
$\Sigma_\k^V(i\w_n)=T\sum_{\q,m} V_{\k,\q}(i\w_n,i\w_m)G_\q(i\w_m)$
vanishes for $\w_n\rightarrow0$  since eq. (\ref{eqn:V})
is a separate function of $\w_n$ and $\w_m$ \cite{Allen}.
%due to the separable interaction in eq. (\ref{eqn:V}).
%Since $V_{\k,\q}(i\w_n,i\w_m)$ in eq. (\ref{eqn:V})
%is a separate function of $\w_n$ and $\w_m$,
%mass-enhancement due to the energy derivative of the 
%normal self-energy $\Sigma_\k(i\w_n)=\sum_{\q,m} V_{\k,\q}(i\w_n,i\w_m)G(i\w_m)$
%vanishes for $\w_n\rightarrow0$ \cite{Allen}.

\begin{figure}[!htb]
\includegraphics[width=.75\linewidth]{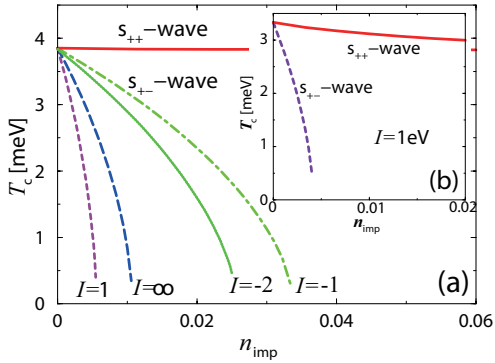}
\caption{(color online)
Obtained $T_{\rm c}$ for the $s_\pm$-wave and $s_{++}$-wave states
as functions of $n_{\rm imp}$.
(a) $g_{1,2}=2$ eV ($T_{\rm c0}=46$ K) and (b) $g_1=3g_2=3$ eV ($T_{\rm c0}=40$ K).
}
\label{fig:Tc}
\end{figure}

To consider the superconductivity due to interband interaction 
between FS1,2 and FS3,4, we put 
$g_{\a,\b}(=g_{\b,\a})=-g_1$ for $(\a,\b)=(1,3),(1,4)$,
$g_{\a,\b}=-g_2$ for $(\a,\b)=(2,3),(2,4)$, and $g_{\a,\b}=0$ for others.
When $\w_D=0.03$, % and $I=\infty$, 
the obtained $T_{\rm c}$'s are shown in Fig. \ref{fig:Tc} for 
(a) $g_1=g_2=2$ eV ($T_{\rm c0}=46$ K at $n_{\rm imp}=0$) and 
(b) $g_1=3g_2=3$ eV ($T_{\rm c0}=40$ K).
%Note that $T_{\rm c0}$ is equal for $s_\pm$- and $s_{++}$-wave states 
%since we neglect the intraband interaction.
In the present choice of $\phi_\k$,
$T_{\rm c0}$ is equal for both $s_\pm$- and $s_{++}$-wave states.
In case (a), the obtained 
$|\Delta_\a|$ in the band diagonal basis
is almost isotropic and the same for $\a=1\sim4$:
In contrast, $|\Delta_1| : |\Delta_2| : |\Delta_{3(4)}|\approx 3:1:1.5$ 
in case (b).
%that the $s_\pm$-wave state in the five-orbital model is very fragile.
In the BCS theory, $-\Delta T_{\rm c} \equiv T_{\rm c0}-T_{\rm c}$ 
is independent of $T_{\rm c0}$, 
and a qualitative relation $-\Delta T_{\rm c}\sim\gamma_\a$ holds
when the Anderson's theorem is violated
\cite{Senga,Allen}.
Since $\gamma_\a$ takes the largest value at $I\sim 1$ eV \cite{comment},
$T_{\rm c}$ for the $s_\pm$-wave state vanishes
at $n_{\rm imp} \sim 0.005$ for $I=+1$ eV. % in Fig. \ref{fig:Tc}.
In the $s_{++}$-wave state,
$T_{\rm c}$ in (b) slowly decreases with $n_{\rm imp}$ with downward convex, 
since weak pair breaking occurs unless $\Delta_\a = \Delta_\b$.
%since the interband scattering induces the weak
%pair-breaking if $\Delta_\a \ne \Delta_\b$.

%indicating the violation of the Anderson's theorem.
%$-\Delta T_{\rm c}$ reaches 40 K/\% for $I=1$ eV in fig. \ref{fig:Tc}.

%In calculating Fig. \ref{fig:Tc}, however, we have taken the 
%$\e_\k$-integration in eqs. (\ref{eqn:Tmat}) and (\ref{eqn:Eliash1})
%analytically by neglecting the particle-hole asymmetry \cite{Allen}.
%Due to this simplification,
%the obtained $T_{\rm c}$ for $I=1,2$ eV in Fig. \ref{fig:Tc}
%is supposed to be {\it overestimated}.
%In the orbital-less model,
%$-\Delta T_{\rm c}=n_{\rm imp}/16N(0)$ K for $n_{\rm imp}\rightarrow0$,
%which is consistent with 
%Similar result is obtained if we use the RPA interaction
%given in Ref. \cite{Kuroki} instead of eq. (\ref{eqn:V).
%We stress that $T_{\rm c}$ is independent of $n_{\rm imp}$ 
%in the orbital-less model in the unitary limit \cite{Senga}.
%The present result manifests the significant role of 
%orbital degree of freedom in multiband SC state.
%In this case, $T_{\rm c}$ for $s_{++}$-wave state is
%almost independent of $n_{\rm imp}$.

%%%%%%%%%%%%%%%%%%%%%%%%%%%%%%%%%
%%%%%%%%%%%%%%%%%%%%%%%%%%%%%%%%%

% renormalization factor
Until now, we have neglected the mass-enhancement,
%factor due to many-body effect,
which is $z^{-1}=m^*/m\approx 2$ by ARPES \cite{ARPES1,ARPES4}.
%and the specific heat measurement \cite{Sato-private2}.
%In this case, the coupling constant $\lambda$ is renormalized to $z\lambda$,
%and therefore $T_{\rm c0}\approx \w_D e^{-1/z\lambda}$ \cite{Allen}.
%(Note that $z^{-1}=1+\lambda$ in the weak-coupling theory.)
%in the electron-phonon interaction.)
Since the depairing effect is renormalized by $z$, 
$-\Delta T_{\rm c}$ is reduced to
$-\Delta T_{\rm c}\cdot z \ (\sim z\gamma_\a)$
\cite{Allen}.
When $z^{-1}=2$, $T_{\rm c}$ for $s_\pm$-wave state vanishes
at $n_{\rm imp}\approx0.01$, $0.02$, and $0.066$ for $I=1$ eV,
$I=\infty$, and $I=-1$ eV, respectively.
Although $-\Delta T_{\rm c}$ ranges from $46$ K/\% to $7$ K/\%,
$T_{\rm c}$  for $s_\pm$-wave vanishes when 
$\rho_{\rm imp}\approx 20 \ \mu\Omega{\rm cm}$ for any $I$
However, Sato et al. studied the impurity effects in polycrystalline 
La1111 and Nd1111, and found that $T_{\rm c}$ vanishes when 
$\rho_{\rm imp}\sim 3 \ {\rm m}\Omega{\rm cm}$,
which corresponds to $\sim 750 \ \mu\Omega{\rm cm}$ in single crystal
 \cite{Sato-imp}.
Also, $\rho$ in single-crystal Fe(Se,Te)
just above $T_{\rm c}=15$ K exceeds 450 $\mu\Omega$cm in ref. \cite{Fe11}, 
mainly due to elastic scattering.
Then, the estimated mean-free-path $l_{\rm mfp}$ is as short as 
$a_{\rm Fe\mbox{-}Fe}=2.8 {\buildrel _{\circ} \over {\mathrm{A}}}$ \cite{comment3}
that is about $1/10$ of the coherence length in Ba122 \cite{coherence}.
(Note that $\rho_{\rm imp}\propto l_{\rm mfp}^{-1}$ is independent of $z$.)
When $l_{\rm mfp}\sim a_{\rm Fe\mbox{-}Fe}$, 
even $s_{++}$-state will be broken by localization \cite{Sato-imp}.

% discussion
In summary, we have studied the effect of 
Fe-site substitution in iron pnictide superconductors.
%on the $s_\pm$-wave and $s_{++}$-wave states.
Due to the presence of orbital degree of freedom,
the $s_\pm$-wave state is as fragile as nodal gap states 
against nonmagnetic impurities.
%due to the interband scattering.
%For example, only 1\% impurities with $I=+1$eV
%is expected to induce a sizable reduction in $T_{\rm c}$; 
%$-\Delta T_{\rm c}\sim 40$K.
The critical residual resistivity for vanishing $T_{\rm c}\sim40$ K
for $s_\pm$-wave state is only $\rho_{\rm imp}^{\rm cr}\sim 20\ \mu\Omega$cm.
The corresponding mean-free-path is $\sim25 a_{\rm Fe\mbox{-}Fe}$,
which is longer than the experimental coherence length.
% $\sim10a_{\rm Fe\mbox{-}Fe}$.

%We have revealed that only 1\% nonmagnetic impurities with $I\ge1$ eV,
%which induces $\rho_{\rm imp}\gtrsim10\ \mu\Omega$cm,
%gives rise to the prominent pair-breaking in the $s_\pm$-wave state.
%This fact leads to the large in-gap states and the great reduction 
%in $T_{\rm c}$.

%promoted by orbital degree of freedom.
%as far as the transformation matrix ${\hat U}_\k$ depends on $\k$.
%Thus, previous analysis of unitary impurity effect
%based on orbital-less models are erroneous for iron pnictides.
%This fact demonstrates the significance of the $\k$-dependence of
%the unitary matrix ${\hat U}_\k$ for the impurity interferential effect.
%originates from the fact that the
%the orbital degree of freedom promotes the 
%interband pair-breaking scattering.
%The obtained result well (but accidentally)
%corresponds to the special case det$\{{\hat I}^{\rm b}\}=0$
%in widely-used orbital-less model in the unitary regime \cite{Senga}.
%The present study illustrates the considerable significance of 
%orbital degree of freedom
%in studying the impurity effect in the SC states. % with sign-reversals.
%Considering the robustness of superconductivity against impurities
%in iron pnictides, it will be an important future issue to find a
%possible mechanism for the $s_{++}$ state.

%cond-mat
%According to this analysis,
%robustness of superconductivity against impurities
%in iron pnictides supports the $s_{++}$-wave state.
%submit
Considering the robustness of superconductivity against impurities
or randomness, the $s_{++}$ state would be a promising candidate 
for iron pnictide superconductors. 
%%%%%%%%%
However, $s_\pm$-wave state will become stable when (i) $|I|\ll1$ eV, 
or the potential radius is comparable to the lattice spacing
%much larger than 1 ${\buildrel _{\circ} \over {\mathrm{A}}}$
and the large momentum scattering is suppressed.
To reveal this possibility, we need more systematic first principle 
calculations for impurity potentials or measurements of $\rho_{\rm imp}$.
Also, the $s_\pm$-wave state can be stable when 
(ii) the $d$-orbital weight on the FS is completely
modified by many-body effect, or 
(iii) strong coupling SC state like in heavily under-doped
high-$T_{\rm c}$ cuprates is realized.
%(e.g., $\xi/a_{\rm Fe\mbox{-}Fe}\sim O(1)$)
%$z^{-1}\gtrsim O(10)$) 
Thus, it is important to study the many-body electronic states
to clarify these possibilities.

\acknowledgements
We are grateful to M. Sato, Y. Kobayashi, Y. Matsuda, T. Shibauchi,
D.S. Hirashima, K. Ueda, K. Yamada, H. Ikeda, T. Nomura, K. Izawa, 
Y. Senga, A.F. Kemper, and co-authors of 
Ref. \cite{Kuroki} for valuable discussions. 
This study has been supported by Grants-in-Aid for Scientific 
Research from MEXT of Japan, and by JST, TRIP.

%%%%%%%%%%%%%%%%%%%%%%%%
%references
%%%%%%%%%%%%%%%%%%%%%%%%


\begin{thebibliography}{99}
\bibitem{comment2}
The formal solution of eq. (\ref{eqn:Tb}) can be expressed as
${\hat T}_{u,v}^{\rm b}
 = \sum_w ({\hat 1}-{\hat I}^{\rm b}{\hat G}^{\rm b})_{u,w}^{-1}
 {\hat I}_{w,v}^{\rm b}$,
where $u=(\a,\k)$ and so on.
Since ${\rm det}\{ {\hat I}^{\rm b}\} =0$ in this representation
in the present model, ${\hat T}_{u,v}^{\bf b}$ is not band-diagonal
even for $I=\infty$.
\bibitem{comment}
In a single band model, 
$\gamma/n_{\rm imp}={\rm Im}T(-i0)
= (1+2a(\pi N(0))^{-2} I^{-1})/\pi N(0) + O(I^{-2})$
for $a\equiv {\rm Re}\{g_{\rm loc}(0)\} \ll N(0)$, which depends on 
the sign of $I$ when $a\ne0$ (particle-hole asymmetric case).
We find that $a>0$ in the present model.
\bibitem{comment3}
Using the $T$-matrix approximation, we numerically derived the relation
$\rho_{\rm imp}\approx 500\times(a_{\rm Fe\mbox{-}Fe}/l_{\rm mfp})\ \mu\Omega{\rm cm}$,
where $l_{\rm mfp}$ is the averaged mean-free-path.
It is also obtained based on the 2D electron gas model;
%by assuming a free electron dispersion; 
the conductivity due to four FSs is $4\times (e^2/h)k_{\rm F}l_{\rm mfp}$
and $k_{\rm F}\sim0.2\pi/a_{\rm Fe\mbox{-}Fe}$.
\end{thebibliography}
\end{document}